# Heterogeneous treatment effect estimation with high-dimensional data in public policy evaluation – an application to the conditioning of cash transfers in Morocco using causal machine learning

Author 1: Patrick Rehill[1]*

Author 2: Nicholas Biddle[1]

[1] Centre for Social Research and Methods, Australian National University

*Corresponding author: patrick.rehill@anu.edu.au

Abstract

Causal machine learning methods can be used to search for treatment effect heterogeneity in high-dimensional datasets even where we lack a strong enough theoretical framework to select variables or make parametric assumptions about data. This paper uses causal machine learning methods to estimate heterogeneous treatment effects in the case of an experimental study carried out in Morocco which evaluated the effect of conditionalizing a cash transfer program on several outcomes including maths test scores which is the focus of this work. We explore treatment effects across a dataset of 1936 pre-treatment variables. For the most part, heterogeneity is modelled by two different factors, participation in education (at the baseline) and more general measures of poverty. Those who are more disadvantaged at the baseline benefit less from any treatment. While conditioning generally has a negative effect this more disadvantaged group is also hurt more by conditioning. The second purpose of this paper is to demonstrate and reflect upon a causal machine learning approach to policy evaluation. We propose a novel causal tree method for interpretable modelling of causal effects and reflect on the difficulty of explaining atheoretical results.

> Policy Significance Statement
>
> There are two different ways in which this work is meant to be significant to policy. The first is in substantive findings. We replicate the average treatment effect findings of the original paper but build upon them with heterogeneous treatment effect estimates. We compare treatment against control and conditioning against labelling. We find a number of variables that drive treatment effect heterogeneity. Across all analyses those with greater participation in education at the baseline and lower general measures of poverty benefit more than those with more disadvantage.
>
> The second implication is for policy evaluation. We show the viability of the causal forest for atheoretical evaluation with high-dimensional data and propose a method called the Distilled Doubly Robust Causal Trees to find clusters of treatment effect heterogeneity.

**1. Introduction**
Conditional Cash Transfer (CCT) programs have become a popular method to address poverty in middle-income countries (Fiszbein and Schady 2009; Cruz et al. 2017; Baird et al. 2011). They address immediate needs by providing cash transfers but also try to incentivise virtuous activity to address deprivation in the long run. However, these programs are not uncontroversial as the conditioning leads to practical questions about whether the additional costs of enforcing compliance are worth the gain (Baird et al. 2011). These costs are borne by the funder who has to pay for the exercise of verifying eligibility, but also (generally in a time cost) by recipients of payments or service providers (like teachers or doctors) who carry the administrative burden of collecting data to prove eligibility (Herd and Moynihan 2018). There are also concerns that those who need the cash most are least likely to comply with conditioning meaning CCTs can have a regressive effect (Heinrich and Knowles 2020).

The Tayssir program in Morocco (Benhassine et al. 2015) was designed to study a novel solution to this problem in a field experiment by comparing a CCT conditional on school attendance against a Labelled Cash Transfer (LCT), a middle-ground between conditional and unconditional policies. The LCT simply gave a payment that was labelled as one to improve school attendance and mandated that enrolment in the program would be through local school principals but there was no enforcement. The hypothesis was that just through this labelling the program will give many of the benefits of a CCT, without the additional cost in time, money and autonomy of a CCT program. In an evaluation of the program, Benhassine et al. (2015) found that receiving any of the treatments (conditional or labelled compared with the control group) had substantial effects on school attendance outcomes, but for the most part it made almost no difference whether the payment was conditioned or labelled.

In this paper we build upon the original evaluation by Benhassine et al. to examine heterogeneity in treatment effects in a more comprehensive way; looking particularly at the outcome of maths test scores. While the original paper looked at some heterogeneous relationships – mostly at the gender of the student and the gender of the recipient of the payment – this paper expands out that analysis to look at the effect of many different possible drivers of heterogeneity. There are two related reasons this may be useful. The first is that capturing heterogeneity in a model helps in making sure there are no sources of heterogeneity that were missed in the original paper's surprising finding, this heterogeneity could potentially be interesting from a policy point of view; the second is that any new sources of heterogeneity could shed further light on the mechanism which makes the LCT roughly as or more effective than the CCT. This paper adopts more flexible methods to search exhaustively for drivers of treatment effect heterogeneity. It uses a causal machine learning method called the causal forest to non-parametrically estimate individual-level conditional average treatment effects (CATEs) which can be explored and aggregated up into group average treatment effects (GATEs) (Wager and Athey 2018; Athey et al. 2018). We then distil this model to a single tree for interpretability. This analysis is intended both to build upon the substantive findings of the original paper and also to demonstrate how the causal forest can simplify analysis of heterogeneous treatment effects in applied work where there are multiple treatments being considered and where there may be complex patterns of heterogeneity.

We mirror the findings of the original evaluation showing treatment has a measurable benefit and the LCT is no worse than the CCT, in fact it seems to give slightly better outcomes. For the most part, the exploration of heterogeneity shows that those with more disadvantage at the baseline benefit less from the treatment overall and in particular suffer more for conditioning.

While this paper intends to make a substantive contribution to the development economics literature, it also intends to make a methodological contribution. The causal forest is still a relatively nascent method and in particular, no existing peer-reviewed study to the best of our knowledge uses a data-set as high-dimensional as the one we use for our analysis here. This is despite the fact that in theory causal machine learning methods allow for the exploration of datasets with many more variables than traditional methods for econometric analysis (Athey and Imbens 2015). This paper demonstrates how such large datasets can be useful where high-dimensional data is available while also grappling with some of the challenges that come with such analysis (namely difficulties posed by correlated features and those in interpreting effects where high-dimensional data and atheoretical analysis makes this challenging). It proposes a novel variant of the causal tree (Athey and Imbens 2016) called a Distilled Doubly Robust Causal Tree to sort through high-dimensional data and model treatment effect heterogeneity in a flexible and interpretable way.

## 2. Literature review
### 2.1. Conditional, unconditional and labelled cash transfers
Conditional cash transfers were popularised in Latin America in the 1990s and 2000s and have proved popular around the world, particularly in middle-income countries (Fiszbein and Schady 2009; Baird et al. 2011). These programs have also proved a popular topic for econometric evaluation due to the relatively common use of random assignment, and there is a reasonable number of public reports and papers that have been published (e.g. Macours, Schady & Vakis 2012; Soares, Ribas & Osorio 2008; Martins & Monteiro 2016; Levy & Ohls 2010; Millán et al. 2019; Palmer et al. 2022). These programs have all targeted both direct poverty reduction, and some activity incentivised by the conditioning, usually school attendance or preventative healthcare interventions like attending post-natal medical check-ups. For the most part, these studies have judged CCTs to be an effective anti-poverty intervention and also an effective incentive to the behaviours they aim to incentivise.

Conditioning however is controversial, some sources have argued that for reasons of cost-effectiveness or to ensure that those who do not comply still get access to much needed financial assistance, programs should not be conditional (Baird et al. 2011; Robertson et al. 2013; Afzal et al. 2019; Fiszbein and Schady 2009; Heinrich and Knowles 2020). There are two published experimental studies that compare CCTs and unconditional cash transfers (UCTs) and both show mixed results. Baird et al. (2011) shows conditioning a cash transfer in Malawi had a positive effect on some outcomes and a negative effect on others. Akresh et al. (2013) studying a program in Burkina Faso find no significant difference in effect on school enrolment overall, but some additional benefit among girls and children who weren't in school at the start of the study from conditioning. It seems that conditioning in some contexts might be beneficial, but that in other contexts it may be unnecessary or even stop very poor people from receiving cash they desperately need (Loeser et al. 2021).

One way of trying to get the 'best of both worlds' of conditional and unconditional policies might be through LCTs. Here labelling encourages a particular virtuous behaviour without the overhead and restrictiveness of conditioning. Benhassine et al. (2015) showed labelling performing as well as or even better than conditional cash transfers. The only other published experimental evaluation of a labelled cash transfer to the best of our knowledge is Heinrich and Knowles (2020) which showed similar results to Benhassine et al. (2015. They study the effect of a transfer for orphans and vulnerable children in Kenya. In this study there was little difference in outcomes for most of the sample and the poorest parts of the sample showed a negative effect from conditioning due to what the authors call a regressiveness in conditioning – that is the poorest are least likely to be able to comply with conditions and therefore less likely to receive a payment. These papers make an attractive case for labelling. However, this raises the question of why labelling seems to be so effective.

### 2.2. Mechanisms behind the effectiveness (or ineffectiveness) of labelled transfers
From an orthodox economics perspective, the idea that labelling would improve performance over simple unconditional transfers or that labelling could even rival the performance of conditioning seems slightly strange. This section explores some of the theory that might be useful in explaining the strange results we see in the two published LCT evaluations.

### 2.2.1. The flypaper effect
The theoretical motivation behind labelled transfers is the so-called 'flypaper effect'. Here, labelling provides a behavioural nudge towards spending money in the intended way (Hines and Thaler 1995). In the case of the Tayssir, the hope is that the transfer will be spent in support of child wellbeing and particularly education. The idea of this effect is that resources that are acquired in different ways are often in practice spent in different ways despite their fungibility. Research into the 'flypaper effect' originally concerned the stimulatory effect of government spending versus an increase in private incomes which in the macroeconomic theory of the time should have had an equal effect on output but in practice does not (Scott 1952; Bailey and Connolly 1998). The specific discussion of the mechanism causing this nonfungibility came with the rise of public choice theory in the 1970s (Bailey and Connolly 1998). The flypaper effect was cast in more personal terms with the rise of behavioural economics and a seminal paper by Hines and Thaler (1995). The key theoretical breakthrough here was applying the idea of mental accounting – that individuals classify different pools of money in different ways and so spend from each in irrational and distinct ways (Thaler 1985; Kahneman and Tversky 2013). This understanding of the flypaper effect would eventually be combined with another key insight of behavioural economics, 'nudge' interventions, to apply the flypaper effect to household use of transfer payments.

Early empirical papers on this topic all focused on child welfare and these found labelling had mixed effects (Laura Blow et al. 2007; Kooreman 2000; Jacoby 2002). The first substantial evaluation of a labelled benefit was first published in 2011 and looked at the effect of the British Government's labelling of the winter fuel payment. It showed the percentage of the transfer spent on fuel was drastically higher (41%) than the percentage of general income spent in the same way (3%) (Beatty et al. 2014). The Tayssir program then was novel in applying the flypaper effect to an anti-poverty intervention.

### 2.2.2 Motivation crowding
Motivation crowding is an idea from behavioural science which is meant to explain cases where the introduction of a material reward seems to disincentivise a behaviour (Frey and Jegen 2000). This effect is particularly strong in cases where there is a qualitative change in reward (i.e. going from no material reward to the introduction of such a reward) compared with quantitative changes in rewards (i.e. an increase in an existing material reward).

This behaviour of course runs counter to the traditional economic conception of incentives and so in the real-world there is often a push and pull between these two incentive forces (Esteves-Sorenson and Broce 2022). Whether the crowding out effect dominates over the additional extrinsic reward depends greatly on the nature of the intrinsic and extrinsic rewards and the people involved. There is unfortunately little work on the interaction between poverty and motivation crowding.

2.2.3 The effect of transfers on 'nevertakers' dominating the effect on 'compliers'
One of the criticisms of CCTs is that by refusing to give transfers to those that do not comply with program conditions, policymakers are condemning these needy non-compliers to deeper poverty than they would have seen under a UCT (Heinrich and Knowles 2020; Fiszbein and Schady 2009). We can think about CCT compliance by analogy to local average treatment effect (LATE) analysis (Angrist et al. 1996). Essentially, the decision to conditionalize a cash transfer involves making the assumption that the benefits to compliers (who are not for example sending their child to school under unconditional payments but would under conditioning) outweigh the harms done to nevertakers who are being stripped of material assistance. It is worth adding that the effect on nevertakers is generally a regressive one as these are often the poorest of poor people, for example people so reliant on the labour of their children that they cannot spare them to go to school (Heinrich and Knowles 2020). In the case of an LCT, one effect that may cause the labelled transfer to perform well compared to a conditional transfer is that the benefits to nevertakers from not conditionalizing might dominate the benefits to compliers of conditionalizing. For example, better nutrition among nevertakers thanks to transfers might boost outcomes in aggregate more than additional schooling for compliers. Combine this with a behavioural economics effect from labelling and there is the potential for LCTs to perform better than either CCTs or UCTs.

2.2.4. Lack of program understanding
A final explanation discussed by Benhassine et al. (2015) is that a lack of program understanding causes labelling to be more effective because recipients believe that the transfer is conditional and so they act as they would if it were conditional. Not only is this a possible explanation for the kinds of effects we see in that original study, but it may also be a threat to external validity as it would mean that we should expect the effect of labelling to be biased upwards in the context of a time-limited field experiment with several treatment arms all called the Tayssir. If the program were rolled out to the whole population indefinitely, and if only a single arm were rolled out at scale, we might expect understanding to grow and therefore the efficacy of labelling to shrink. To put it simply, if people who understand the program show the CCT being more effective relative to the LCT, we will have reason to be suspicious that the positive LCT results may not hold over time. It is worth noting that we might expect some of this possible confusion to be reduced in the trial thanks to area-based randomisation of treatment.

2.3. Potential drivers of heterogeneity
As the literature specifically on the effect of labelling for intrahousehold allocation of resources is relatively sparse, it is useful to reach into a broader range of literature to understand factors that may drive heterogeneity in the effect of labelling over conditioning. One variable that does have a demonstrated effect in the intra-household flypaper effect literature (albeit only in one paper) is income. Jacoby (2002) argued that the intra-household flypaper effect weakens for poorer households, so the effect of labelling increases as income increases. Another factor that may affect this intrahousehold effect is the number of children in a household. Jacoby identifies the number of children taking part in a school meals program he studies has an income effect on calorie intake within the household. This, coupled with the way in which each child is funded within a household, but at different rates depending on age suggests the number of children may have an effect (possibly a nonlinear one) on treatment in the Tayssir.

It is common in surveys relating to poverty in developing countries to rely a lot on the characteristics of the household head to explain differential outcomes, particularly around gender where a female household head is often taken as a proxy for increased status of girls in the household (Rajaram 2009; Klasen et al. 2011). Since the head of household might be the most powerful decision-maker around schooling decisions, we can take characteristics around them as predictors. An important hypothesis from the original evaluation for instance is this idea that misunderstanding the LCT as conditional could have driven up effects in a way that would not last in the long-term if recipients became more familiar with the program and a CCT under the same name was not running in parallel (Benhassine et al. 2015). The kinds of variables that might lead to a household head misunderstanding the program (for example illiteracy) could be indicative of such an effect.

There may be attributes of the school that children attend that drives treatment effect heterogeneity when it comes to learning outcomes. This is in-line with the literature arguing that supply-side interventions (improving services) not just demand side interventions (incentivising participation) are necessary in contexts quality of services are low (Fiszbein and Schady 2009). Such an effect has been shown specifically in the Tayssir data where increased demand for poorly resourced services can be detrimental to test scores (Gazeaud and Ricard 2024). This is in other words one mechanism which might lead to a regressive effect from the transfer.

Finally, characteristics of students themselves might drive heterogeneous effects. As with many societies around the world, there is a good deal of research showing that girls' education is still not valued as highly as boys' in Morocco (Derdar 2014; Chafai 2017; Ennaji 2016). Benhassine et al. (2015) demonstrates a number of heterogeneities driven by gender in their Tayssir evaluation. There is also good research showing that (as one might expect), school leaving becomes more common as Moroccan students get older, in part due to higher opportunity cost and in part due to barriers to secondary education (EL Alaoui et al. 2021). In addition, the design of the Tayssir program where the cash value increases over time, with hard steps up, suggests there may be interesting non-linear age effects.

## 3. Program background

Morocco is a lower middle income country which at the time of the intervention had relatively high (though declining) levels of poverty (Achy 2011). One particularly worrying outcome was poor education statistics (low adult literacy, low school completion) as poor education can create intergenerational entrenchment of poverty (Benhassine et al. 2015). As is the norm with CCTs, the Tayssir was designed to address this specific cause of poverty and also provide cash aid to alleviate poverty in general.

The Tayssir began in 2008 and roll-out was done in such a way that there was area-level randomisation in treatment allocation to allow for clean causal identification. Some areas were randomised into a control group (735 households), others were given one of four treatments, these are listed below along with the same size for each.
- LCT (Group 1) – 1684 households.
- CCT with monitoring by school teachers (Group 2) – 1309 households.
- CCT with monitoring by school teachers with inspectors to audit attendance as well (Group 3) – 1199 households.
- CCT with monitoring via a biometric system using fingerprint recording to take school attendance (Group 4) – 1217 households.

We study both the effect of treatment over not receiving treatment and also the effect of conditioning over labelling.

## 4. Data and methods
### 4.1. Identification and replication of ATE calculation

The first step in this research is replicating the analysis in the original evaluation but estimating an average treatment effect with double machine learning to address the small imbalances between groups in the original sample that resulted from area-level randomisation (Chernozhukov et al. 2018; Benhassine et al. 2015). All the analysis in this paper will be drawing on a linked dataset compiled from the datasets used in the evaluation paper. While that paper has quite a wide scope, this one will be focusing specifically on one outcome – the results of a mathematics test. The reason for this is that the outcome of primary interest in the original paper was a binary outcome that was relatively rare in the sample (although there is a larger sample size with just administrative data missing most of the covariates which was used in the original paper). This lack of variance in the variable and the small sample for which we have many covariates means that looking for heterogeneity here would be at best fruitless and at worst misleading. The maths test outcome though, while not as crucial to the original evaluation, offers a measure of improvement in education through a variety of mechanisms (increased attendance, better learning due to better nutrition, increased salience of education within the household to name a few). While it is one limited measure, the intent was for this test to act as a reasonable proxy for measuring the effect of the treatment on learning achievement while keeping the test simple enough to be administered by a surveyor (Benhassine et al. 2015).

The allocation of treatment has been randomised by school (and therefore by geographic community), however there are some differences in treatment group characteristics. The original paper used a regression model to model out the effect of the factors that were unbalanced between the treatment and control samples. This paper will use an augmented inverse probability weighting approach (AIPW) to estimate average treatment effect and CATEs. This is the standard approach for causal forests and gives a non-parametric, doubly robust estimate (Robins et al. 1994; Athey et al. 2018).

For all this analysis we use an index of the mathematics questions administered to students as the dependent variable. These cover four categories – digit recognition, number recognition, subtraction and division. While the number of questions vary, we normalise each category to a maximum score of one and sum the results. While the original evaluation does not give specifics on how their index of maths ability was calculated, the results from our approach are similar. The distribution of marks is shown below in Figure 1.

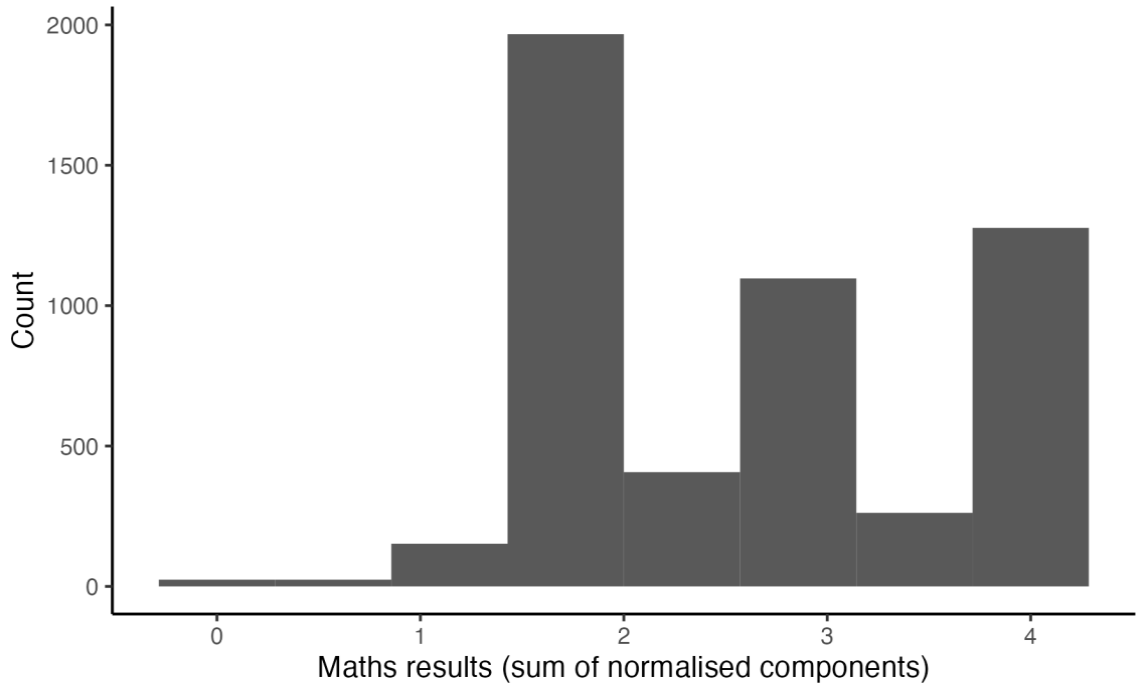

Figure 1: Histogram of maths scores

### 4.2. Extending the research with heterogeneous treatment effect analysis

This paper will extend the original analysis by analysing treatment effect heterogeneity using the causal forest algorithm from the *grf* package (Athey et al. 2018) and a novel causal tree approach. This technique allows for accurate, non-parametric estimation of CATEs. Due to the relative novelty of the method, there is little established methodology built up around it to test causal effects for CATEs (Athey and Wager 2019; Elek and Bíró 2021; Lechner 2019).

One of the ways in which causal machine learning models can be used is to apply a model to a high-dimensional dataset and allow the model to find those variables which give the best estimate of treatment effect (see Chernozhukov et al. (2018) for more details on this approach). This includes every variable measured at the household level, every variable measured at the individual-level (across all household members including the household head) and every variable measured at the child-level (again across all children in the household). We also construct sets of these variables that are specifically for the child who took the maths test. There are 1936 valid, numeric, pre-treatment variables. There are some variables that are arguably ordinal, at least locally on the scale (such as time-use codes) and we cautiously include them in this analysis assuming that if they are not useful, the algorithm will simply ignore them and if they are, we can look at the scale to try and try to interpret the effects.

However, we cannot possibly fathom the effects of all these variables and understanding the effects of variables in a causal model is crucially important if we are to get insights from it (Rehill and Biddle 2023). While the causal forest benefits from high dimensional data purely in terms of fit, this still creates challenges in the interpretation of effects because it does not provide an easy way of knowing which variables explain the treatment effect differences. The problem is fundamentally one of feature selection. The most commonly used solution for feature selection in the causal forest is to use the procedure proposed for predictive forests by Basu et al. (2018) and suggested for use with the causal forest by Athey and Wager's (2019) influential application paper. This approach fits a forest and then refits a forest without any variables that have below mean variable importance. Variable selection is calculated by making a depth-weighted count of the variables used for splitting in the ensemble (Athey and Wager 2019). This approach was attempted however is not used in the final paper. In brief, problem here is that in a setting with relatively weak signal and many different types of data, this procedure seems to favour selecting continuous variables to fit noise well over ordinal variables with short scales that might be important drivers of actual heterogeneity. This is particularly a problem as the random forest randomly samples variables, inflating the number of splits made on worse predictors. This of course improves the prediction accuracy of the ensemble (Breiman 2001), but does make this variable importance measure interpretable.

The solution we use in this paper is a variation on an approach that has been taken elsewhere – fitting a representative tree to help make sense of the structure of the forest. The simplest way to do this is to fit a single causal tree on the raw data (Athey and Imbens 2016) but this is rarely done due to the instability of a tree, instead drawing a tree out of a forest is more common. Different approaches have been suggested for this including finding a tree that is most representative via a tree-distance metric (Hui-min Zhou et al. 2023; Zhuoming Zhou et al. 2023), finding the tree that best minimises the R-Loss objective (Wager 2018) or fitting a causal tree based on variables selected via variable importance metrics in the causal forest (Raghavan et al. 2022). In this paper we will take an approach inspired more by knowledge distillation (Dao et al. 2021; Frosst and Hinton 2017; Domingos 1997), the reason for this is that there is good reason to expect a distilled tree to fit the underlying treatment effect distribution well given it is designed to create unbiased semi-parametric estimates in a way that should improve distillation as well (Dao et al. 2021). In this approach we first fit a causal forest on all variables using 10,000 trees. Then we fit what Athey and Imbens (2016) call an 'honest' tree ala the original causal tree using half the data for fitting and half for estimation. When fitting we are not using the causal tree criterion but instead simply trying to fit the causal forest predictions with a regression tree (knowledge distillation). Then, the estimation step uses the doubly robust scores from the other half of the data to estimate effects for leaves. We can then bootstrap confidence intervals by taking the tree structure as fixed and generating a bootstrap distribution of estimates for each leaf with 2000 replicates. We call this a Distilled Doubly Robust Causal Tree (DDRCT). This is like a cross between the original causal tree and the random-forest-as-adaptive-kernel in the generalised random forest. This creates a tree that should fit better than a standard causal tree that we would get from fitting directly on data or in an ensemble, and it gives asymptotically normally, doubly robust estimates (Athey and Imbens 2016; Athey et al. 2018). However, this tree is still unstable (that is the structure changes greatly with small perturbations in the sample, meaning it is unlikely the tree is fitting an underlying transportable (Bareinboim and Pearl 2012) causal process). To improve stability and therefore the usefulness of the structure in telling us something about the causal mechanism we fit 1000 trees in this manner on subsamples of the data (each a randomly selected 50%) and then select the tree that best fits the distribution of causal forest predictions out of sample. This selects for stability or low variance.

The trees are fit to a depth of 3 but doubly robust estimates are made for each node including nodes that are not terminal. There is currently no principled way to select the ideal depth for a causal tree and in the absence of this, seeing estimates at different depths provides a good way to understand effects. This means that splitting into small subgroups without sufficient sample for credible effect estimation can be ignored.

We fit a forest and then a DDRCT for the effect of (any) treatment over the control, the effect of (any) conditioning over labelling and finally we use a multi-arm causal forest (Athey et al. 2018) to estimate effects for each conditioning treatment separately over the LCT and use the DDRCT to explore the most promising treatment – biometric scanners to enforce conditioning.

## 5. Results

### 5.1. Comparing treatment to control

#### 5.1.1. Average treatment effects

The first part of this analysis compares treatment (that is any cash transfer) against the control. Here we can see that there is a statistically measurable effect at the 95% confidence level. There is also a measurable effect for the labelled transfer and the conditional transfer with biometric enforcement.

Table 1. Average treatment effect estimate comparing different treatments against the control.

| Contrast | Estimate | SE |
| --- | --- | --- |
| All treatments – control | 0.086 | 0.030 |
| Labelled transfer – control | 0.131 | 0.044 |
| Conditional transfer 1 – control | 0.081 | 0.051 |
| Conditional transfer 2 – control | -0.007 | 0.060 |
| Conditional transfer 3 – control | 0.115 | 0.058 |

### 5.1.2 Treatment effect heterogeneity

Looking to treatment effect heterogeneity we plot the distribution of individual treatment effects for all treatments on a histogram in Figure 2. This figure also shows the ATE point estimate and the error distribution around it. The distribution here does not suggest obvious heterogeneity, the estimates are distributed roughly Gaussian, according to the distribution of the average treatment effect. However, we can see if there are any variables driving heterogeneity by looking at the effects of specific variables.

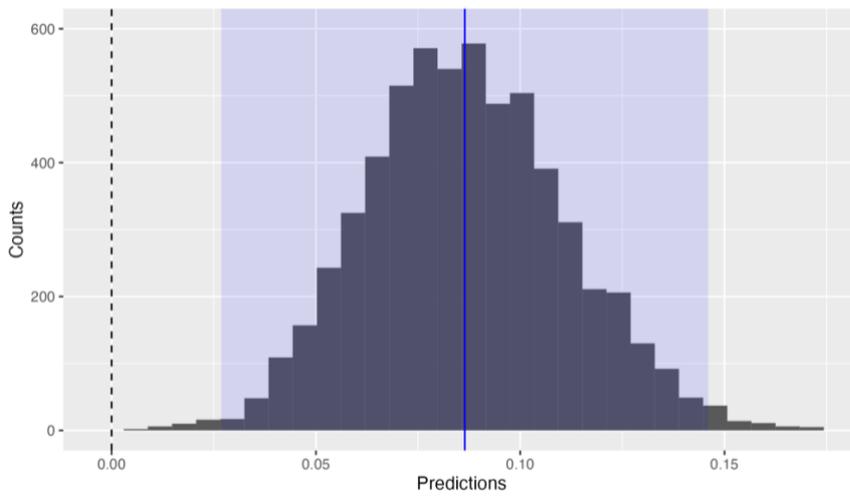

Figure 2: Distribution of causal forest treatment effects for any treatment compared to the control group.

Distilling the original forest into a decision tree and then estimating doubly robust scores for each node we see a couple of things immediately. The first is that the root node is different from the average treatment effect estimate. This is due to the fact this estimate is made only with doubly robust scores from the estimation sample. We can see that the estimates within nodes seem largely in-line with the estimates causal forest estimates, but more dispersed around the average treatment effect (both on the negative and positive sides). It is likely the forest itself has a regularising effect on the individual treatment effect estimates in many different ways (in particular, a more complex kernel function with larger bandwidths). These estimates are however still valid, asymptotically normal (because of honesty and out-of-sample prediction from the causal forest), doubly robust (because of the AIPW estimator) estimates.

The most important variable here is the amount spent on couscous per month which is the variable on which the root splits. Those who spend more than 21 dirhams on couscous the previous month had a substantially higher treatment effect (0.234) than those who did not (0.075). Below this are several statistically significant leaves that are descendants of the this leaf and which have larger effects. The next split with statistically significant results is on the child's principal activity between 17:00-17:30. If this value is below 72, the effect is roughly doubled (to 0.444). The value of 72 is roughly the dividing line between work within the household or education and work outside the household and leisure. Figure 3 graphs the choices in this variable. Choices are aggregated up to the level of a one-digit classification and categorised. It seems likely that what we are seeing here is an effect driven by being in or out of educational activities in this time period at the baseline. This group benefits by far the most of any. This is likely the reason for splitting on time use in other nodes as well.

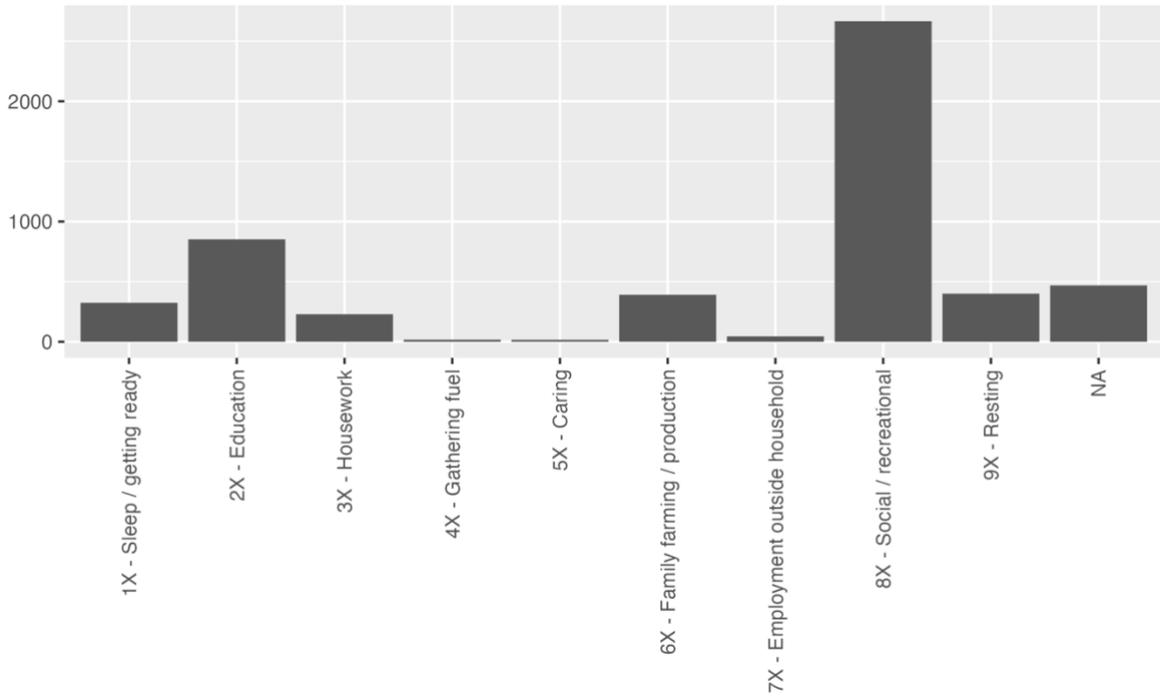

Figure 3: Activities between 17:00 and 17:30

There is also a statistically significant effect for those engaged in activities with a code above 72 during this period (0.216) particularly if they were also in a household spending more than 15.5 dirhams on transport in the past month (0.313).

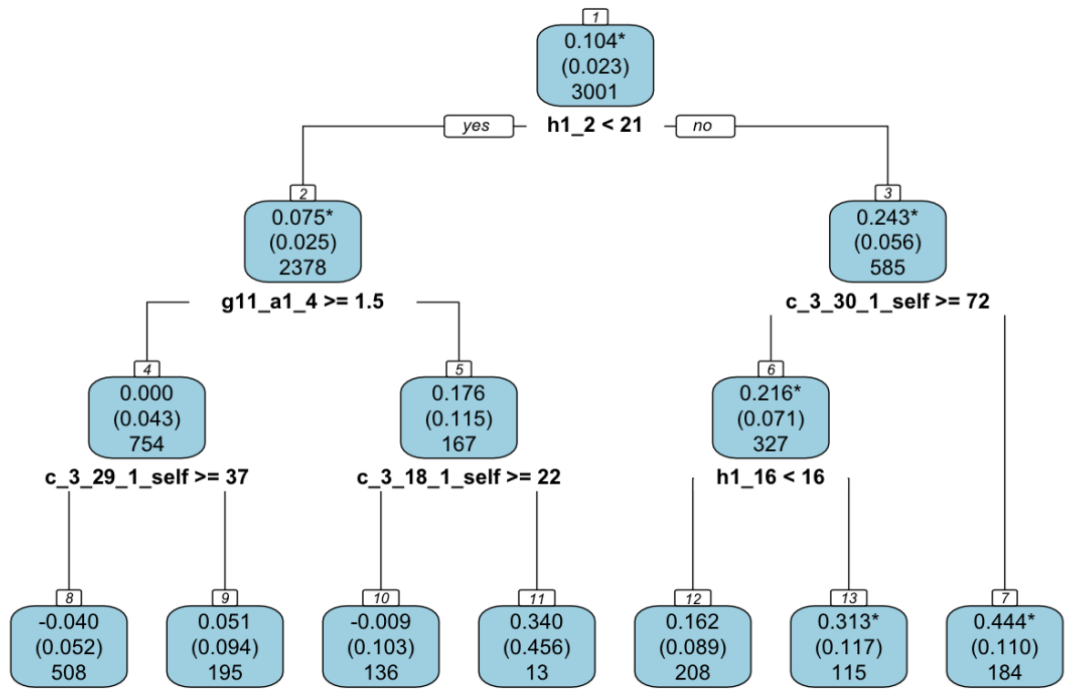

Figure 4: DDRCT for treatment over control

* Effect statistically significant at α = 0.05

Note: h1_2 is spending on couscous in dirhams in the past month. g11_a1_4 is whether house member 4 (often a child) performed an activity to aid the household in the past 12 months, generally employment or farming (1 being yes, 2 being no). c_3_30_1_self is the activity the test taking child was doing at 17:00 – 17:30 the previous day at the baseline survey. c_3_29_1_self is the activity the test taking child was doing at 16:30 – 17:00 the previous day at the baseline survey. c_3_18_1_self is the activity the test taking child was doing at 11:00 – 11:30 the previous day at the baseline survey. h1_16 is spending on transportation in the past month.

Table 2: Nodes from DDRCT for treatment over control with detailed rules.

| Node | Depth | Est | SE | | n | Rules |
|---|---|---|---|---|---|---|
| 1 | 0 | 0.104 | 0.023 | * | 3001 | |
| 2 | 1 | 0.075 | 0.025 | * | 2378 | • Monthly couscous spending < 21 |
| 3 | 1 | 0.243 | 0.056 | * | 585 | • Monthly couscous spending >=21 |
| 4 | 2 | 0.000 | 0.043 | | 754 | • Monthly couscous spending <21<br>• Household member 4 worked in activity 1 in last 12 months = No |
| 5 | 2 | 0.176 | 0.115 | | 167 | • Monthly couscous spending < 21<br>• Household member 4 worked in activity 1 in last 12 months = Yes |
| 6 | 2 | 0.216 | 0.071 | * | 327 | • Monthly couscous spending >=21<br>• Tested child's activity from 17:00 – 17:30 >=72 |
| 7 | 2 | 0.444 | 0.11 | * | 184 | • Monthly couscous spending >=21<br>• Tested child's activity from 17:00 – 17:30 < 72 |
| 8 | 3 | -0.04 | 0.052 | | 508 | • Monthly couscous spending < 21<br>• Household member 4 worked in activity 1 in last 12 months = No<br>• Tested child's activity from 16:30 – 17:00 >=37 |
| 9 | 3 | 0.051 | 0.094 | | 195 | • Monthly couscous spending < 21<br>• Household member 4 worked in activity 1 in last 12 months = No<br>• Tested child's activity from 16:30 – 17:00 < 37 |
| 10 | 3 | -0.009 | 0.103 | | 136 | • Monthly couscous spending < 21<br>• Household member 4 worked in activity 1 in last 12 months = Yes<br>• Tested child's activity from 11:00 – 11:30 >=22 |
| 11 | 3 | 0.34 | 0.456 | | 13 | • Monthly couscous spending < 21 |

| | | | | | |
|---|---|---|---|---|---|
| | | | | | • Household member 4 worked in activity 1 in last 12 months = Yes |
| | | | | | • Tested child's activity from 11:00 – 11:30 < 22 |
| 12 | 3 | 0.162 | 0.089 | | 208 | • Monthly couscous spending >=21 |
| | | | | | • Tested child's activity from 17:00 – 17:30 >=72 |
| | | | | | • Monthly spending on transportation< 16 |
| 13 | 3 | 0.313 | 0.117 | * | 115 | • Monthly couscous spending >=21 |
| | | | | | • Tested child's activity from 17:00 – 17:30 >=72 |
| | | | | | • Monthly spending on transportation>=16 |

### 5.2. Comparing conditional to unconditional treatments

Having contextualised the treatments against the control, the rest of the analysis focuses on comparing LCT with CCT treatments. Table 3 presents analysis for a new causal forest fit to estimate the effect of different types of conditioning when compared against labelling.

Table 3. Average treatment effect estimate comparing individual conditioning treatments against labelling treatment.

| Contrast | Estimate | SE |
|---|---|---|
| All conditioning - labelling | -0.086 | 0.037 |
| Group 2 - labelling | -0.066 | 0.047 |
| Group 3 - labelling | -0.157 | 0.054 |
| Group 4 - labelling | -0.043 | 0.049 |

Overall, the comparison of all conditioning to labelling and the comparison of Group 3 with the LCT shows a statistically significant effect, both negative.

Looking at the distributions of treatment effects we can see two important points. The first is that the effects are roughly normally distributed which might be suggestive of little heterogeneity (though does not necessarily rule it out). We also see a decent portion of each distribution is on the positive side of zero, suggesting there may be small gains from conditionalizing the program for a minority of recipients, however, this may just be noise in estimation.

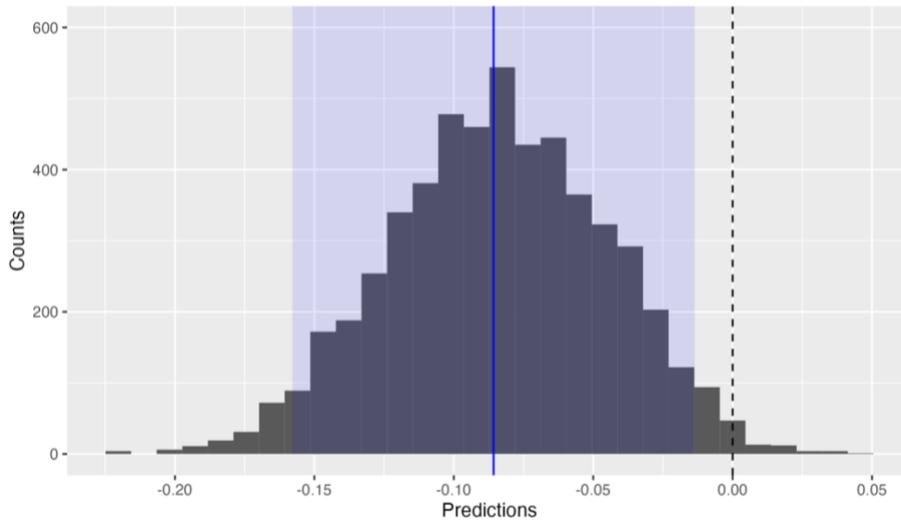

Figure 5: Distribution of treatment effects of labelling versus all conditioned treatments (dashed line signifies zero, blue line is AIPW ATE estimate, shaded region is 95% confidence interval for this analysis).

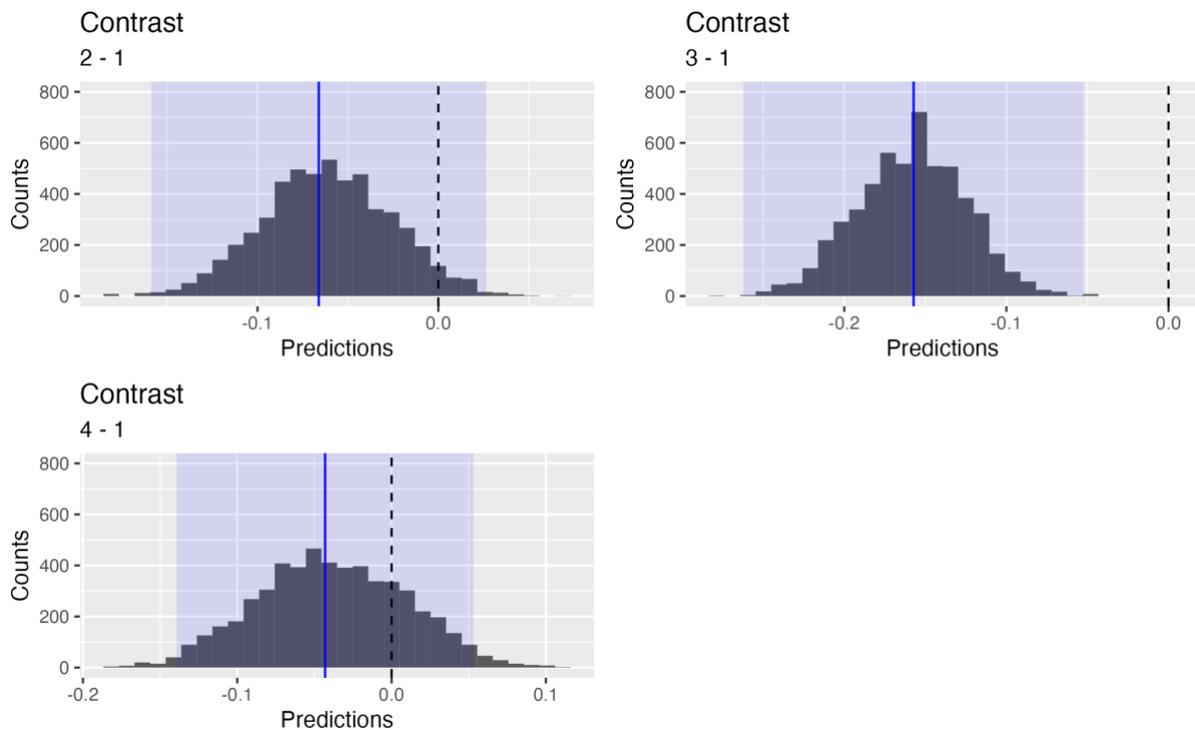

Figure 6: Distribution of treatment effects across all treatments (dashed line signifies zero, blue line is AIPW ATE estimate, shaded region is 95% confidence interval for this analysis).

### 5.3. The effect of individual variables

#### 5.3.1. All conditioning treatments against labelling

As in the original paper, we see a small, negative effect for any conditioning over labelling. This is a theoretically interesting result which was discussed at length in the original paper. Our focus here will instead be on trying to understand the heterogeneity in this effect and if there are subsamples where the treatment effect is different (perhaps different enough for there to be a policy case for narrower conditioning).

The most important variable here is whether there is a third child aged 6 – 15 in the household at the time of enumeration. There is a relatively obvious reason this might be useful on face value – it tells us if a lot of school-age children are present at time of enumeration – but the fact this is collected is also interesting. There is no variable for two children present or four children present. Is there some use of this question that affects program administration or data collection. As far as we can see there is no reason for this provided in any English-language material on the Tayssir.

There is also a significant effect for Child 1 (generally the eldest child at home) being currently enrolled in school. If they are not enrolled in school at the baseline, the effect of conditioning over labelling is worse. If Child 2 is also currently not enrolled in school, the magnitude of this (negative) effect is even greater.

The last node with a statistically significant effect is actually a positive one. The node only contains 54 individuals in the estimation sample. These people are the group where there was not a third child home during enumeration, where there was only one child between 6 and 15 in the household altogether and where the house roof material was "Other".

Note for the variables in this dataset, children are indexed with different numbers within the household as are household members so a child could be child 1 and household member 4. This does not necessarily mean Child 1 is the eldest although it seems that there tends to be a birth order in child indexing.

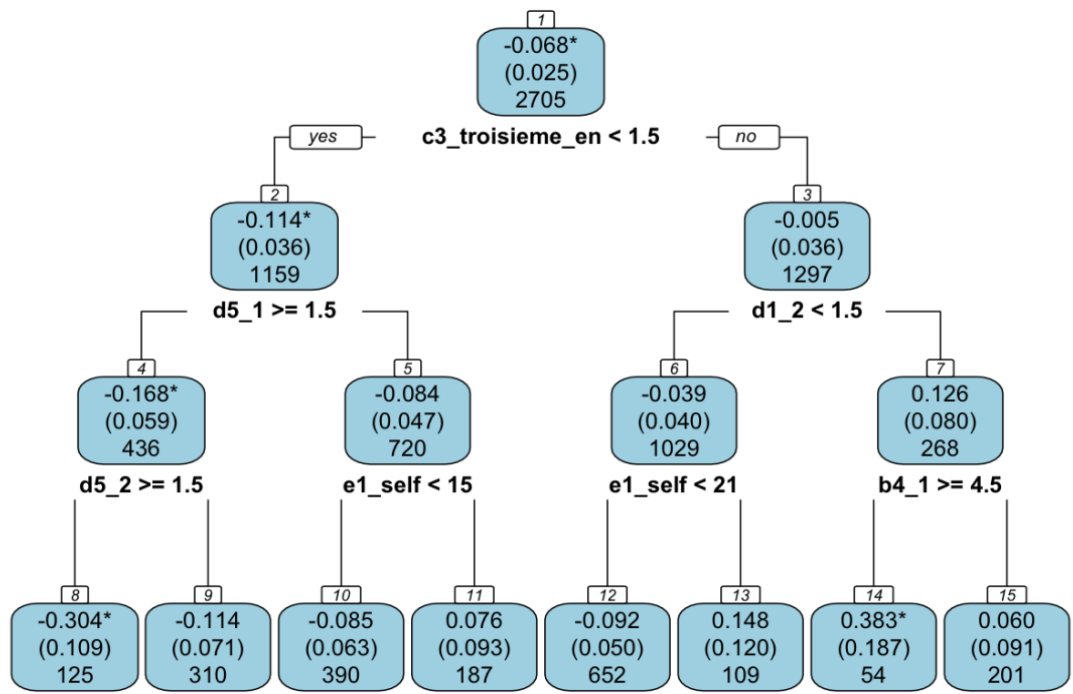

Figure 7: DDRCT for all conditioning against labelling

* Effect statistically significant at $\alpha = 0.05$
Note: c3_troisieme_en is whether a third child aged 6 – 15 is present at the baseline (1 being yes, 2 being no). d5_1 is whether child 1 is currently enrolled in school (1 being yes, 2 being no). d1_2 is whether there is a second child in the household (1 being yes, 2 being no). d2_self is the test-taking child's ID within the household. d5_2 is whether child 2 is currently enrolled in school (1 being yes, 2 being no). e1_self is the amount spend on school fees for the test-taking child in the past month (at baseline) b4_1 is the roof material of the dwelling, >= 4.5 corresponds with a roof material of 'other'.

Table 4: Nodes from DDRCT for all conditioning against labelling with detailed rules.

| Node | Depth | Est | SE | | n | Rules |
|---|---|---|---|---|---|---|
| 1 | 0 | -0.068 | 0.025 | * | 2705 | |
| 2 | 1 | -0.114 | 0.036 | * | 1159 | • Third school-aged child present = Yes |
| 3 | 1 | -0.005 | 0.036 | | 1297 | • Third school-aged child present = No |
| 4 | 2 | -0.168 | 0.059 | * | 436 | • Third school-aged child present = Yes <br> • Child 1 currently enrolled in school = No |
| 5 | 2 | -0.084 | 0.047 | | 720 | • Third school-aged child present = Yes <br> • Child 1 currently enrolled in school = Yes |
| 6 | 2 | -0.039 | 0.04 | | 1029 | • Third school-aged child present = No <br> • Another child between 6 and 15 in household = Yes |
| 7 | 2 | 0.126 | 0.08 | | 268 | • Third school-aged child present = No <br> • Another child between 6 and 15 in household = No |
| 8 | 3 | -0.304 | 0.109 | * | 125 | • Third school-aged child present = Yes <br> • Child 1 currently enrolled in school = No <br> • Child 2 currently enrolled in school = No |
| 9 | 3 | -0.114 | 0.071 | | 310 | • Third school-aged child present = Yes <br> • Child 1 currently enrolled in school = No <br> • Child 2 currently enrolled in school = Yes |
| 10 | 3 | -0.085 | 0.063 | | 390 | • Third school-aged child present = Yes <br> • Child 1 currently enrolled in school = Yes <br> • Amount spent on school fees for test taking child < 14.5 |
| 11 | 3 | 0.076 | 0.093 | | 187 | • Third school-aged child present = Yes <br> • Child 1 currently enrolled in school = Yes <br> • Amount spent on school fees for test taking child >=14.5 |
| 12 | 3 | -0.092 | 0.05 | | 652 | • Third school-aged child present = No <br> • Another child between 6 and 15 in household = Yes <br> • Amount spent on school fees for test taking child < 21 |
| 13 | 3 | 0.148 | 0.12 | | 109 | • Third school-aged child present = No <br> • Another child between 6 and 15 in household = Yes <br> • Amount spent on school fees for test taking child >=21 |

| | | | | | |
|---|---|---|---|---|---|
| | | | | | • Third school-aged child present = No |
| | | | | | • Another child between 6 and 15 in household = No |
| 14 | 3 | 0.383 | 0.187 | * 54 | • Roof material = "Other" |
| | | | | | • Third school-aged child present = No |
| | | | | | • Another child between 6 and 15 in household = Yes |
| 15 | 3 | 0.06 | 0.091 | 201 | • Roof material = not "Other" |

### 5.3.2. The best case for conditioning – Biometric enforcement

Rather than looking through each treatment separately, we will simply look at the only conditioning treatment that seems to rival labelling – enforcing attendance with biometric scanners. This is of course ignoring the cost, practical and possibly ethical issues that this treatment poses over labelling (Innovations for Poverty Action and Abdul Latif Jameel Poverty Action Lab 2009; Benhassine et al. 2015). Is there any subgroup that benefits more from this treatment than simple labelling and is this group large enough to be policy-relevant? This is the group that had biometric scanning as the enforcement mechanism. Interestingly enough, due to rollout problems, for most of this group for most of the study period the treatment was in effect a labelled transfer as the scanners were inoperable (Innovations for Poverty Action and Abdul Latif Jameel Poverty Action Lab 2009).

The results for this treatment are interesting as the variables here seem to relate mostly to the size of household whether than be number of children (different numbers of children being in the household) or adults (first child household ID as children are numbered after adults).

As expected, there is no statistically significant effect at the root node, however once splitting reaches depth 2 we start to see statistically significant effects in both directions. For households with a third school age child present and a fourth child of school age in the household there is a statistically significant negative effect of -0.225 (as compared to no measurable effect if there is not a fourth child). The subset of this group where Child 1 is not enrolled in school (likely a proxy for any school enrolment, or the enrolment of older children) the effect is even more negative (-0.352). For those without a third school-aged child present and without a second school-aged child in the household, there is a statistically significant positive effect for conditioning of 0.283. For those with a third child present at enumeration, no fourth school-aged child in the household and a small overall household size (where the child taking the test was household ID 4 or less) there was a strong positive effect estimated of 0.379 while those where the household ID was greater than 4 had a negative effect of -0.185.

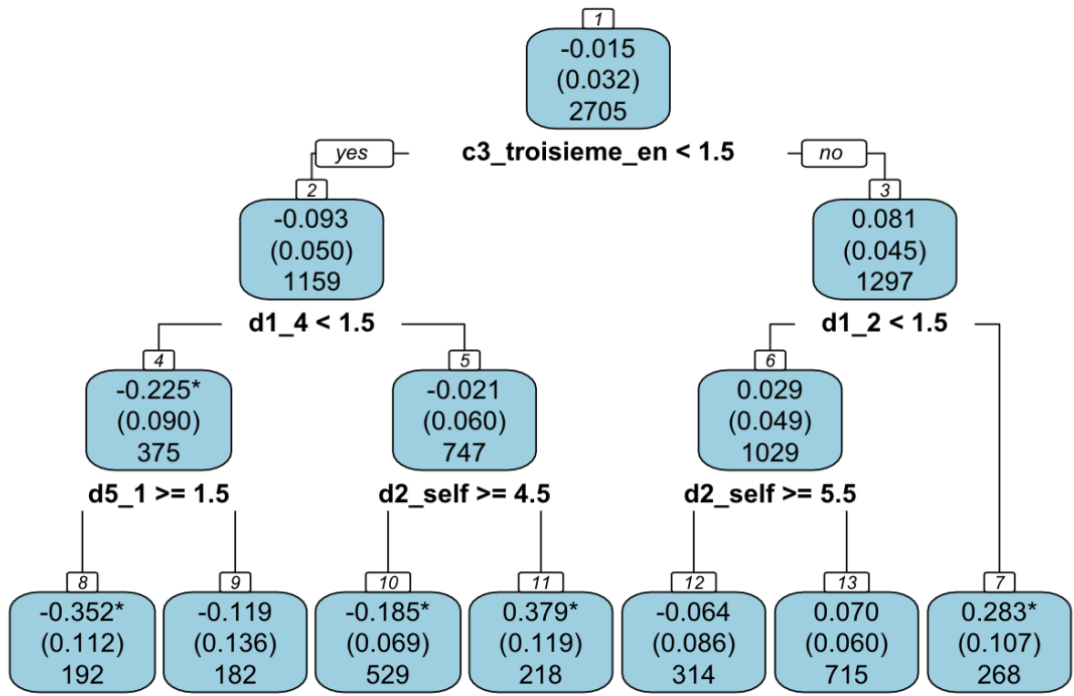

Figure 8: DDRCT for biometric conditioning
* Effect statistically significant at $\alpha = 0.05$
Note: c3_troisieme_en is whether a third child aged 6 – 15 is present at the baseline (1 being yes, 2 being no). d1_4 is whether there is a fourth child in the household (1 being yes, 2 being no). d1_2 is whether there is a second child in the household (1 being yes, 2 being no). d5_1 is whether child 1 is currently enrolled in school (1 being yes, 2 being no). d2_self is the test-taking child's ID within the household.

Table 5: Nodes from DDRCT for biometric conditioning with detailed rules.

| Node | Depth | Est | SE | | n | Rules |
|---|---|---|---|---|---|---|
| 1 | 0 | -0.015 | 0.032 | | 2705 | |
| 2 | 1 | -0.093 | 0.05 | | 1159 | • Third school-aged child present = Yes |
| 3 | 1 | 0.081 | 0.045 | | 1297 | • Third school-aged child present = No |
| 4 | 2 | -0.225 | 0.09 | * | 375 | • Third school-aged child present = Yes<br>• Fourth child between 6 and 15 in household = Yes |
| 5 | 2 | -0.021 | 0.06 | | 747 | • Third school-aged child present = Yes<br>• Fourth child between 6 and 15 in household = No |
| 6 | 2 | 0.029 | 0.049 | | 1029 | • Third school-aged child present = No<br>• Second child between 6 and 15 in household = Yes |
| 7 | 2 | 0.283 | 0.107 | * | 268 | • Third school-aged child present = No<br>• Second child between 6 and 15 in household = No |

| | | | | | |
|---|---|---|---|---|---|
| 8 | 3 | -0.352 | 0.112 | * 192 | • Third school-aged child present = Yes |
| | | | | | • Fourth child between 6 and 15 in household = Yes |
| | | | | | • First child currently enrolled in school = No |
| 9 | 3 | -0.119 | 0.136 | 182 | • Third school-aged child present = Yes |
| | | | | | • Fourth child between 6 and 15 in household = Yes |
| | | | | | • First child currently enrolled in school = No |
| 10 | 3 | -0.185 | 0.069 | * 529 | • Third school-aged child present = Yes |
| | | | | | • Fourth child between 6 and 15 in household = No |
| | | | | | • First child household ID >=4.5 |
| 11 | 3 | 0.379 | 0.119 | * 218 | • Third school-aged child present = Yes |
| | | | | | • Fourth child between 6 and 15 in household = No |
| | | | | | • First child household ID < 4.5 |
| 12 | 3 | -0.064 | 0.086 | 314 | • Third school-aged child present = No |
| | | | | | • Second child between 6 and 15 in household = Yes |
| | | | | | • First child household ID >=5.5 |
| 13 | 3 | 0.07 | 0.06 | 715 | • Third school-aged child present = No |
| | | | | | • Second child between 6 and 15 in household = Yes |
| | | | | | • First child household ID < 5.5 |

## 6. Discussion
### 6.1. Explaining results in treating versus control
When it comes to looking at individual variables, there is much room for interpretation. Those who spend more on couscous at the baseline see a substantially higher benefit than those who spend less. This may be because those that spend more are in bigger households, those who are getting their food through the formal economy as opposed to farming for their own use, or those who are not suffering from actual food poverty. All plausible factors affecting the efficacy of a payment (Fiszbein and Schady 2009; Baird et al. 2011; Heinrich and Knowles 2020). The other variables that are important here are taking part in education-related activities at certain hours in the baseline survey. This is a good measure of enrolment and participation in general, however, the choice of hours need the end of the working day suggest that homework, taking part in school activities or longer school days may be particular causal mechanisms here.

It is interesting that those who are more deprived benefit less from the treatment even when simply comparing a cash payment with no cash payment. We might expect the gains to be highest for those where the payment represents a large chunk of household income, potentially being enough to pull a household out of absolute poverty. But this is not in fact what we see. There are a few possible reasons for this. The first is that the comparison is not just between an automatic payment and no payment. It is between a raft of payments, most in the sample being on conditioned payments and no payment. Households that cannot or do not comply with these conditions lose their benefits and so, these poorer households being less likely to comply might be expected to receive fewer transfers on average. This reflects Heinrich and Knowles' (2020) ideas about the regressiveness of conditioning and our idea of 'nevertakers'. In addition, there may be some measurement sensitivity problems here. It may be that for very poor children, the marginal increase in their ability to do maths from additional schooling and additional resources at home (including learning materials but also more secure food, easier access to safe water and all the other multidimensional depravations that can affect learning) is simply not enough to get them to a level where they can improve their scores on the maths test. Finally, the transfer is a relatively small amount compared to similar programs elsewhere meaning that even for the very poor, the extrinsic incentive effect of the payment may be low and the intrinsic effect of labelling or conditioning dominates (Benhassine et al. 2015). In line with the regressiveness argument, we might expect less disadvantaged households to be more sensitive to these intrinsic effects. We might expect the flypaper effect in particular to be weaker among these poorer households (Jacoby 2002).

### 6.2. Explaining results in conditioning versus labelling

The most important finding from this analysis is that there was minimal heterogeneity in outcomes, and as with the original study, the effect of conditioning was largely non-existent or negative. This is a very important substantive outcome as it suggests that in this case it is possible to have a best of both worlds between UCT and CCT with an LCT. One can get outcomes equal to or even better than the CCT while maintaining the low overheads and lack of regressive effects of a UCT (Heinrich and Knowles 2020; Benhassine et al. 2015). However, there was still some heterogeneity in results.

The presence of a third school-aged child at home during the enumeration was important here. If there was a child home effects of conditioning were measurably negative while if they was not the effect was not measurable. The importance of school enrolment at baseline here is interesting too. It says that the conditioning is actually most effective (or perhaps least harmful) among households where students are already in education. Whatever incentive effect there is to enrol is cancelled out by some other factor. This heterogeneity strengthens the case for there being a regressive effect (Heinrich and Knowles 2020). That is, that the loss of payments for those who do not meet the conditions creates more harm than the incentive effect of the conditioning creates benefit. This is likely an economically regressive effect because it is likely those worst off who will be least likely to meet the conditions.

While we have good theory on why labelling might yield similar results to conditioning, it is harder to justify why it might be measurably more effective than conditioning, particularly when conditioning has no statistically significant effect over even the control. Of course, the effect sizes involved are small, it may simply be sampling error or small errors in randomisation from group randomisation that cannot be modelled out by the causal forest.

### 6.3. Explaining results in biometric conditioning versus labelling

This set of results is particularly interesting because while we do see some role for school enrolment here, the heterogeneity is driven above all by the size of household in terms of number of children and number of adults. In general, conditioning is more effective for households that are smaller on both measures. There are broad patterns across many contexts where poverty correlates both with family size (Wietzke 2020) and educational outcomes and participation (Engle and Black 2008). Specifically, a larger family size might weaken the flypaper effect (Jacoby 2002). In addition, the income effect of the payment may be less in a larger family with more adults, non-school-aged children or younger children who attract smaller payments.

There could also be program administration reasons for this effect. It is unclear from the Tayssir documentation if the conditionality is enforced across all children or if the payment for each child is conditionalised separately. If there are more children who all have to be at school, we might expect that is a more stringent conditioning in practice.

### 6.4. Methodological reflections

The difficulty in trying to explain these results shows a drawback in these methods – that the method can easily estimate effects that cannot be easily explained within existing theoretical frameworks. When we are working atheoretically it can become difficult to explain effects regardless of the size and structure of the data. One solution to this problem would be a mixed-methods approach. This data collection would not just be useful in building support for findings, but also in refuting findings that may not actually reflect underlying causal relationships building a more holistic picture of causation (Johnson et al. 2019) (though not at the expense of some philosophical inconsistencies around the place of data-driven theory-building in causation that have yet to be satisfyingly resolved (Pearl 2021)). This would mean in effect adopting an explanatory mixed-methods approach where we use the causal machine learning methods to generate results and then explain those results with qualitative research (Guetterman et al. 2019; Plano Clark and Ivankova 2022). This makes the exercise of using a heterogeneous treatment effect learner an abductive policy evaluation exercise in the vein described by Levin-Rozalis (2000). Another approach is to generate theoretically valid hypotheses either from the causal machine learning model or through this mixed-methods approach and then test them deductively with standard quantitative methods. We can already speculate about some possible ways to test hypotheses generated from this paper. What is the underlying causal mechanism correlating spending on couscous with treatment effect for example? Is it larger households? Richer households? Reliance on subsistence farming? Food insecurity? We could construct a model projecting these variables onto the doubly robust scores to separate out their effects from each other and test hypotheses (Semenova and Chernozhukov 2021; Tibshirani et al. 2021).

High dimensional data can also (as it does in this case) bring along with it many correlated features. This is a challenge because it can be hard to sift through many correlated variables, trying to determine what the underlying causal mechanisms are. This should be a particular problem when using variable importance to reduce the dimensionality of the data (Bénard and Josse 2023; Hines et al. 2022). The causal tree approach proposed here should hopefully address some of these issues (by being inherently low-dimensional in its model, by not relying on random sampling of features like trees within a random forest and by prioritising stability by assessing many candidate trees), however this is not a solved problem. There could be many possible solutions to this problem, one that jumps to mind is to do some kind of nonparametric factor analysis in order to discover an underlying factor structure from the data (or from the forest model). While newer variable importance measures that are less heuristic than counting splits are promising and could prompt new dimensionality reduction and variable selection approaches, they are still computationally expensive for high-dimensional data (Bénard and Josse 2023; Hines et al. 2022).

## 7. Conclusion

This analysis used causal machine learning methods to attempt to find heterogeneity in the Tayssir cash transfer study. We find important heterogeneity in the effect of a treatment over the control case however, for the most part conditioning was not beneficial over labelling for anyone and heterogeneity was mostly a matter of less or more harm. In both cases we saw the benefit of the treatment was greater (or the harm less) for those who were less deprived at the baseline in terms of education and also in terms of general material poverty suggesting that at least when it comes to education outcomes, the policy will not necessarily reduce inequality in outcomes. A labelled transfer (or possibly just an unconditional one) will be the most effective treatment for helping the most impoverished people in this sample.

The downside of the flexible search for treatment effect heterogeneity is that we lack explanations that are as convincing as those that might come from smaller, more theoretically driven models. These may of course miss some effects particularly where programs are idiosyncratic with a lack of a strong theoretical framework readily available (as is the case here with this being the first evaluation of its kind on labelled transfers). The causal forest and DDRCT have in effect generated and tested data-driven hypotheses in a way that avoids p-hacking through sample splitting (Athey and Imbens 2016), however, the existing theory does not provide all the explanations needed to interpret these hypotheses as actual causal effects or alternatively to rebuke some of these findings as essentially trivial correlations. More work is needed to make these methods a practical part of a policy evaluation pipeline. Taking this approach as a jumping off point for the testing of specific hypotheses or exploratory qualitative research is likely to be particularly useful.

**Provenance.** This article was submitted for consideration for the 2024 Data for Policy Conference to be published in Data & Policy on the strength of the Conference review process.
**Funding statement.** None
**Competing interests.** None
**Data availability statement.** Replication data and code can be found in Harvard Dataverse: https://doi.org/10.7910/DVN/29014. Code is available from https://github.com/pbrehill/tayssir_causalforest.
**Author contributions.** Patrick Rehill: Conceptualisation, Methodology, Writing – Original draft, Writing- Reviewing and Editing, Investigation. Nicholas Biddle: Supervision, Writing- Reviewing and Editing